\documentclass[letterpaper]{article}

\usepackage{natbib,alifeconf}  
\usepackage{amsmath,amssymb,amsfonts}
\usepackage{algorithmic}
\usepackage{algorithm}
\usepackage{graphicx}
\usepackage{textcomp}
\usepackage{xcolor}
\usepackage{subcaption}
\usepackage{soul}

\newcommand{\bx}{\mathbf{x}}
\newcommand{\bv}{\mathbf{v}}
\newcommand{\bp}{\mathbf{p}}

\newcommand{\IfThen}[2]{\algorithmicif\ #1\ \algorithmicthen\ #2}
\newcommand{\ElifThen}[2]{\algorithmicelse \algorithmicif\ #1 \algorithmicthen\ #2}
\newcommand{\Else}[1]{\algorithmicelse\ #1}

\title{Tracking Multiple Fast Targets With Swarms: \protect\\ Interplay Between Social Interaction and Agent Memory}
\author{Hian Lee Kwa$^{1,2}$, Jabez Leong Kit$^{1}$ \and Roland Bouffanais$^3$ \\
\mbox{}\\
$^1$Singapore University of Technology and Design, Singapore,\\ $^2$Thales Solutions Asia, Singapore,\\ $^3$University of Ottawa, Canada \\
hianlee\_kwa@mymail.sutd.edu.sg} 

\begin{document}
\maketitle

\begin{abstract}

The task of searching for and tracking of multiple targets is a challenging one. 
However, most works in this area do not consider evasive targets that move faster than the agents comprising the multi-robot system. This is due to the assumption that the movement patterns of such targets, combined with their excessive speed, would make the task nearly impossible to accomplish. In this work, we show that this is not the case and we propose a decentralized search and tracking strategy in which the level of exploration and exploitation carried out by the swarm is adjustable. By tuning a swarm's exploration and exploitation dynamics, we demonstrate that there exists an optimal balance between the level of exploration and exploitation performed. This optimum maximizes its tracking performance and changes depending on the number of targets and the targets' movement profiles. We also show that the use of agent-based memory is critical in enabling the tracking of an evasive target. The obtained simulation results are validated through experimental tests with a decentralized swarm of six robots tracking a virtual fast-moving target.

\end{abstract}

\section{INTRODUCTION}

Swarming multi-robot systems (MRS) are gaining increasing amounts of attention as they bring several advantages compared to their centrally controlled MRS counterparts. This includes system flexibility---the ability to operate in dynamic environments, robustness---the ability to cope with individual agent failures, and scalability---the ability to carry out tasks in systems comprised of different number of agents. As such, swarming MRS have been demonstrated to be able to carry out several tasks such as dynamic area monitoring~\citep{Zoss2018}, area mapping~\citep{Kit2019}, and target monitoring~\citep{Coquet2019}.

In the field of target tracking, the task of monitoring multiple moving targets is an NP-hard problem and was first formalized in the Cooperative Multi-robot Observation of Multiple Moving Targets (CMOMMT) framework by \cite{Parker1997}. Under this framework, agents position themselves to maximize the amount of time that each target is observed. In this study, the authors developed a basic algorithm where robots switched between a `search' and `track' mode. While in the `search' mode, robots repelled each other to disperse and search for the targets. Upon encountering a target, a robot changed to the `track' mode and autonomously moved to the center of mass of all observed targets and its neighbors using a locally calculated force vector. Later, \cite{Parker2002} assigned weights to the targets, preventing target coverage overlap by reducing the influence of targets already under observation on the force vector.

When the number of targets exceeds the number of agents, \cite{Kolling2006} developed a strategy based on the targets' speed and direction of travel. Agents broadcast a `help request' if it predicted that a target was about to move out of detection range. \cite{Esterle2017} studied the implementation of various response models and communications strategies for situations where the number of targets in the search space was known \textit{a priori}. They demonstrated that increasing the level of inter-agent communication served to improve the system-level monitoring performance, regardless of the response model used. 

However, \cite{Mateo2017} have shown that increasing levels of connectivity causes a reduction in an MRS's capacity to respond to dynamic stimuli. Furthermore, \cite{Mateo2019} established that a swarm must change its level of connectivity to maximize its response, adapting itself according to the speed of evolution of the environment. Applying this concept to the tracking of a single fast-moving non-evasive target, \cite{Kwa2020a} showed that an optimum level of connectivity occurs at which the ideal balance between the amount of exploration and exploitation is carried out, thus maximizing the tracking performance of the system. A similar observation was also made by~\cite{Hamann2018}. In this work, a stick pulling task is carried out by a MRS swarm in which the degree of collaboration among agents is varied by changing an agent's sensor range. The author concluded that the amount of information needs to be moderated to maximize a system's performance. Doing so prevent agents from trying to optimize the problem independently, which occurs at low levels of connectivity, and also prevents all agents from solving the same problem in parallel.

In CMOMMT, all proposed strategies are limited to having targets slower than or at the same speed of the individual swarming agents. This is especially the case when evasive targets are being tracked where there is a long-standing assumption that the targets will always be able to evade their pursuers given their superior mobility and maneuverability. This gives a false sense that the problem is impossible for the pursuing swarm~\citep{Parker1997, Parker2002}. However, it has been shown that this is possible in various target capture games where the pursuers have vision of the entire environment~\citep{Janosov2017, Zhang2019}. \cite{Shishika2019} also demonstrated that information sharing between swarming agents aids agents in the process of intercepting and capturing a fast-moving evader. Furthermore, \cite{Ni2011} showed that the networks used to share information among agents must be a dynamic one. This allows connections between agents to be broken and established, maximizing the system's ability to capture evasive targets.

Our contribution to this challenging problem is a novel swarm-based strategy for memory-enabled agents. Memory is demonstrated to be critical in allowing an MRS to track evasive targets that move faster than the individual units. Its introduction also gives another parameter that can tune an MRS's exploration and exploitation dynamics (EED), permitting agents to successfully prioritize either exploratory or exploitative actions. In doing so, the system autonomously adapts its collective dynamics to maximize its performance while tracking different target numbers, speeds, and movement profiles. In this problem of dynamic target tracking problem, \cite{Jordehi2014} cited two main challenges which are addressed in this work, namely: (1) the trade-off between exploration and exploitation carried out by a swarm, and (2) problems associated with outdated system memory. As such, we also thoroughly explore the intricate interplay between these two factors that give rise to the optimum amount of system engagement necessary to maximize the system's tracking performance.

The rest of this paper is structured as follows. We first present the MRS search and tracking strategy. We then describe the operating conditions of the search environment in which we deploy our MRS, as well as introduce a metric that allows for the quantification of a swarm's EED. This novel approach is thoroughly analyzed through simulations, and also validated experimentally using a testbed comprising six decentralized miniature robots.

\section{Methods}

\subsection{Search and Tracking Strategy}
The strategy used by \cite{Kwa2020a} was composed of two regulated behavioral patterns: (1) promotion of agent aggregation around a point of attraction (exploitation), and (2) an adaptive inter-agent repulsion behavior (exploration). These behaviors, inspired by the Charged Particle Swarm Optimization algorithm (CPSO)~\citep{Blackwell2002} and the Social-Only PSO~\citep{Engelbrecht2010}, generated two velocity vectors at each time-step that were combined to give a final agent velocity vector:
\begin{equation}
    \bv_i[t] = \bv_{i,\text{att}}[t] + \bv_{i,\text{rep}}[t],
    \label{eqn:movement}
\end{equation}
where $\bv_{i,\text{att}}[t]$ and and $\bv_{i,\text{rep}}[t]$ are the velocity vectors generated by the attractive component and the repulsion component respectively. Selecting the degree, $k$, of the inter-connecting $k$-nearest neighbor communications network controlled the amount of social interaction between the swarming agents, and hence the overall EED of the swarm.

The overall strategy employed in the system is detailed in Algorithm~\ref{alg:strategy}. The individual components of this algorithm will be explained in the following sections.

\begin{algorithm}
    \caption{: Dynamic $k$-Nearest Network Search and Tracking Strategy}
    \label{alg:strategy}
    \begin{algorithmic}[1]
    \STATE Set $t = 0$, $k \in [2, N-1]$, $\omega=1$, and $ c=0.5$
    \WHILE{System active}
        \FOR{All agents $i \in [1, N]$}
            \STATE Set $\bp$ using Algorithm~\ref{alg:set_p}
            \STATE Calculate $\bv_{\text{att},i}$ using Eq.~\eqref{eqn:vel_update}
            \STATE Calculate $\bv_{\text{rep},i}$ using Algorithm~\ref{alg:adaptive_repulsion}
            \STATE $\bv_i[t] \gets \bv_{\text{att},i}[t] + \bv_{\text{rep},i}[t]$
            \STATE $\bv_i[t] \gets (v_{\text{max}}/v_i[t]) \cdot \bv_i[t]$
            \STATE $\bx_i[t+1] \gets \bx_i[t] + \bv_i[t]$
        \ENDFOR
        \STATE $t \gets t+1$
    \ENDWHILE
    \end{algorithmic}
\end{algorithm}

\subsubsection{Agent Aggregation}
The aggregation component of the strategy was used to generate a point of attraction for the agents, encouraging exploitative actions. Each agent keeps track of the position at which a target was found and the time at which it was detected. Each agent also receives a set of target positions and encounter times from its $k$-nearest neighbors. These received values are compared to an agent's own values and the most recent target position is used as a point of attraction, $\bp$ (see Algorithm~\ref{alg:set_p}). Therefore, agents can exploit both information that is directly sensed from the environment and those coming from its neighbors. Note that a neighborhood is understood in the network sense; an agent $i$ has as many neighbors as its degree $k$. Given that time-varying network topologies are considered, the neighborhoods of each individual agent change over time.

Using an agent's previous velocity and location relative to $\bp$,
an agent's velocity is updated according to:
\begin{align}
    \bv_{i, \text{att}}[t+1] &= \omega \bv_i[t] + c r \big(\bp[t+1] - \bx_i[t+1]\big).
    \label{eqn:vel_update}
\end{align}
This equation is similar to that used in the social-only PSO model~\citep{Engelbrecht2010}, where $\omega$ is the velocity inertial weight, set at $\omega=2$, $c$ is the social weight, set at $c=2$, and $r$ is a number randomly drawn from the unit interval. In computational optimization, this is the main driver of a the system's exploitative behavior. Here, it is used to drive the MRS towards the target. It should be noted that the targets will never overlap each other and that agents do not assign unique identifiers to tracked targets.

Crucial to the system's ability to track an evasive target is the implementation of agent-based memory. It was previously determined that the use of memory was counterproductive as its usage resulted in the exploitation of outdated information, causing swarm aggregation in a location at which the target is no longer present~\citep{Coquet2019, Kwa2020a}. Despite these disadvantages, memory usage has been shown to encourage the aggregation of agents around high quality target patches in static non-destructive foraging tasks~\citep{Falcon-Cortes2019, Nauta2020b}. In the pursuit of an evasive target that moves faster than any individual agent, the use of agent-based memory gives the swarm a longer lasting point of attraction. This increases the amount of exploitation carried out by the MRS, allowing it to close in on a target even though agents are unable to detect the presence of the target. As such, each agent is given a memory, $M$, with a duration of $t_\text{mem}$. 

\begin{algorithm}
\caption{: Point of Attraction Update Algorithm}
\label{alg:set_p}
    \begin{algorithmic}
    
    \STATE Initialize $M = t_\text{mem}$
    
    \IF{Agent detects target}
        \STATE $\bp_{\text{self}} \gets$ Target's position
        \STATE $t_{\text{best}} \gets t$
    \ENDIF
    
    \STATE Determine $\mathcal{N}_i = \{j \in [1, N]$ s.t. agent $j$ is a topological \textit{k}-nearest neighbor of agent $i$\}
    
    \STATE Get list of all neighbors' $\bp$ and $t_{\text{best}}$
    \STATE $\bp_{\text{neigh}} \gets \text{Most recent entry in all neighbors' } \bp$
    \STATE $t_{\text{neigh}} \gets \text{Most recent entry in all neighbors' } t_{\text{best}}$ 
    
    \STATE \IfThen{$t_{\text{best}} + M < t$}{$\bp_{\text{self}} \gets \emptyset$}
    \STATE \IfThen{$t_{\text{neigh}} + M < t$}{$\bp_{\text{neigh}} \gets \emptyset$}
    \STATE \IfThen{$\bp_{\text{self}} = \emptyset \textbf{ and } \bp_{\text{neigh}} = \emptyset$}{$\bp[t] \gets \bx_i[t]$}
    \STATE \ElifThen{$t_{\text{best}} > t_{\text{neigh}}$}{$\bp[t] \gets \bp_{\text{self}}$}
    \STATE \Else{$\bp[t] \gets \bp_{\text{neigh}}$}
    
    \end{algorithmic}
\end{algorithm}

\subsubsection{Adaptive Repulsion}
\label{sec:adaptive_repulsion}
The implemented adaptive repulsion behavior was used to prevent the agents from flocking within a small area and promote area exploration. From a practical robotics standpoint, this behavior also offers an anti-collision measure as a direct byproduct of this mechanism. This behavior was first introduced in \cite{Kwa2020a} and is summarized in Algorithm~\ref{alg:adaptive_repulsion}.

Each agent, $i$, with a set of topological neighbors, $j$, calculates its velocity vector as follows: 
\begin{equation}
    \bv_{\text{rep},i}[t] = - \sum_{j\in \mathcal{N}_i}\left( \frac{a_R[t]}{r_{ij}[t]}\right)^d \frac{{\mathbf{r}_{ij}[t]}}{r_{ij}[t]},
    \label{eqn:rep}
\end{equation}
where $\mathbf{r}_{ij}[t]$ is the vector from agent $i$ to agent $j$ at time-step $t$. The level of inter-agent repulsion is controlled by two factors: (1) the dynamic repulsion strength, $a_R[t]$, controlling agent separation when the system is in equilibrium and, (2) the exponential term $d$ in the pre-factor term $(a_R/r_{ij})$.

The key aspect of this repulsion scheme is the ability of each agent to tailor its repulsion strength, $a_R[t]$, based on information gathered through direct measurements taken from the environment and from communications with its neighbors. Here, we introduce the concept of an agent's tracking state, $S_i[t]$. When an agent has information of a target's location, it is assigned a tracking state of 1, moves towards the target location, and reduces its $a_R[t]$ value until a minimum value is reached. When an agent has no target information, it enters an exploratory state, assigns itself a tracking state of 0, and increases its $a_R[t]$ value until a maximum value is attained. Formally, the tracking state is assigned by considering $S_i(\bx_i[t],t)=0$ if $\bp_i[t] = \emptyset$, and $1$ otherwise.

\begin{algorithm}
    \caption{: Adaptive Repulsion}
    \label{alg:adaptive_repulsion}
    \begin{algorithmic}
    \STATE Set $a_{R,\text{min}} = 2$, $a_{R,\text{max}} = 12$, $d=6$, $\delta_{\text{explore}}=0.1$ and $\delta_{\text{track}}=0.75$
    \WHILE{System active}
        \STATE \IfThen{$a_R > a_{R, \text{min}}$ \AND $S_i[t] = 1$}{$a_R \gets a_R - \delta_{\text{track}}$}
        \STATE \algorithmicelse \algorithmicif{ $a_R< a_{R, \text{max}}$ \AND $S_i[t] = 0$} \algorithmicthen
        \STATE \hspace{\algorithmicindent} $a_R \gets a_R + \delta_{\text{explore}}$
    \STATE Calculate $\bv_{\text{rep},i}$ using \eqref{eqn:rep}
    \ENDWHILE
    \end{algorithmic}
\end{algorithm}

\subsubsection{Swarm Communications Network}
\label{sec:network}
The swarm communication network regulates the EED of the MRS. Adjustment of the level of connectivity, also known as the degree connectivity, of an MRS can have considerable effects on the collective dynamics of the swarm~\citep{Mateo2017,Mateo2019}. \cite{Kwa2020a, Kwa2020b} also established that in the tracking of fast-moving targets that lower degrees of connectivity favor exploration of the domain while higher degrees of connectivity favor domain exploitation.

\subsection{Target Representation}
\label{sec:target_rep}
In this work, the targets do not emit a gradient field; agents are either able to detect a target if they are within a target's radius or are completely unable to do so should they be positioned otherwise. The absence of a gradient field makes the tracking problem more challenging by eliminating the possibility of using gradient-descent methods. This conservative approach represents one of the most challenging cases with a near-zero-range sensor tracking a target faster than the agent themselves. It is only through the deployment of an MRS in such challenging scenarios that it is able to fully make use of its swarm intelligence.

The targets move according to `non-evasive' and `evasive' policies. Using the non-evasive policy, the targets move towards random waypoints within the search space. Using the evasive policy, the targets initially follow the non-evasive policy until it makes contact with an agent. Upon contact with an agent (i.e. when an agent falls within the target's radius), the target calculates its velocity using the repulsion equation presented in~\eqref{eqn:rep}, with all agents within its radius used as repulsion neighbors. After encountering agents for $t_{\text{limit}}$ consecutive iterations, the target travels in a straight line for $t_{\text{evade}}$ time-steps to attempt to outrun its pursuers.

\subsection{Problem Statement}
In this work, a set of tracking agents ${A=\{a_1, a_2, \ldots, a_N\}}$ and a set of targets ${O=\{o_1, o_2, \ldots, o_M\}}$  move within a bounded two-dimensional square search space of dimensions $L \times L$ devoid of any obstacles. Both agents and targets have an $x$ and $y$ position, ${\bx_i = (x_i, y_i)}$, and maximum velocities of $\bv_{a, \text{max}}$ and $\bv_{o, \text{max}}$ respectively, where ${\bv_{a, \text{max}} \leq \bv_{o, \text{max}}}$. The targets are modeled using disc-shaped binary objective functions with fixed radii of ${\rho = L/25}$. A target is considered to be tracked if an agent lies within its radius. Formally: 
\begin{equation}
\label{eqn:coverage}
    \text{cov}(o_m, t)=
    \begin{cases}
        1 & \exists i \in A \text{ s.t. } \|\bx_i - \bx_m\| \leq \rho, \\
        0 & \text{otherwise.}
    \end{cases}
\end{equation}
The goal of the system is to maximize its tracking performance of the targets within the environment, given by the reward function:
\begin{equation}
\label{eqn:cost_fn}
    \Xi = \frac{1}{T J} \sum^T_{t=1}\sum^M_{m=1}\text{cov}(o_m, t),
\end{equation}
where $T$ it the total time period of interest and $M$ is the total number of targets within the search space. In the simulations performed, the agents are tasked with tracking the targets in an environment free from obstacles and are assumed to have perfect information about the target's location once within the target's radius.

\subsection{Exploration and Exploitation Dynamics}
In addition to the tracking performance of the swarm, the collective dynamics of a swarm can be studied through the quantification of the EED of the system. Previously, this was performed by finding the correlation between an agent's direction of travel and bearing to the target~\citep{Kwa2020a,Kwa2020b}. However, a new metric needs to be considered since multiple targets are being studied. As such, a swarm's \textit{engagement ratio} is used to quantify its EED. An agent is considered to be engaged with a target if it has entered the `tracking' state and $S_i(\bx_i[t], t) = 1$. Therefore, the overall engagement ratio of the swarm is calculated as follows:
\begin{equation}
    \label{eqn:engagement}
    \Theta = \frac{1}{NT}\sum^T_{t=1}\sum^N_{i=1} S_i[t],
\end{equation}
where $N$ is the total number of agents within the swarm. With this metric, a higher engagement ratio will indicate that agents are spending a larger proportion of time attempting to track a target, and hence higher levels of exploitation. Conversely, at lower engagement ratios, agents spend a lower proportion of time attempting to track a target, which is characteristic of higher levels of exploration.

\section{Simulation Results}

A homogeneous swarm of $N=50$ agents with a maximum speed of $0.1$ arbitrary-distance-units per time-step was initialized in a $L\times L$ square operating environment, with $L=25$. The agents are tasked with tracking $J$ targets following either an evasive or non-evasive motion policy, where $J \in \{ 1, 2, 3 \}$. By fixing the agents' maximum speed at $v_{a_,\text{max}} = 0.1$ we studied the effects of changing the maximum speed of the targets relative to the agents. All simulations were also carried out over a period of 100,000 iterations, resulting in low variability between runs with different random seeds (below 1\% for both tracking performance and system engagement).

\subsection{Impact of Varying Degree of Connectivity}
\subsubsection{Single Target Tracking}

\begin{figure}[htbp]
    \centering
    \includegraphics[width=0.45\textwidth]{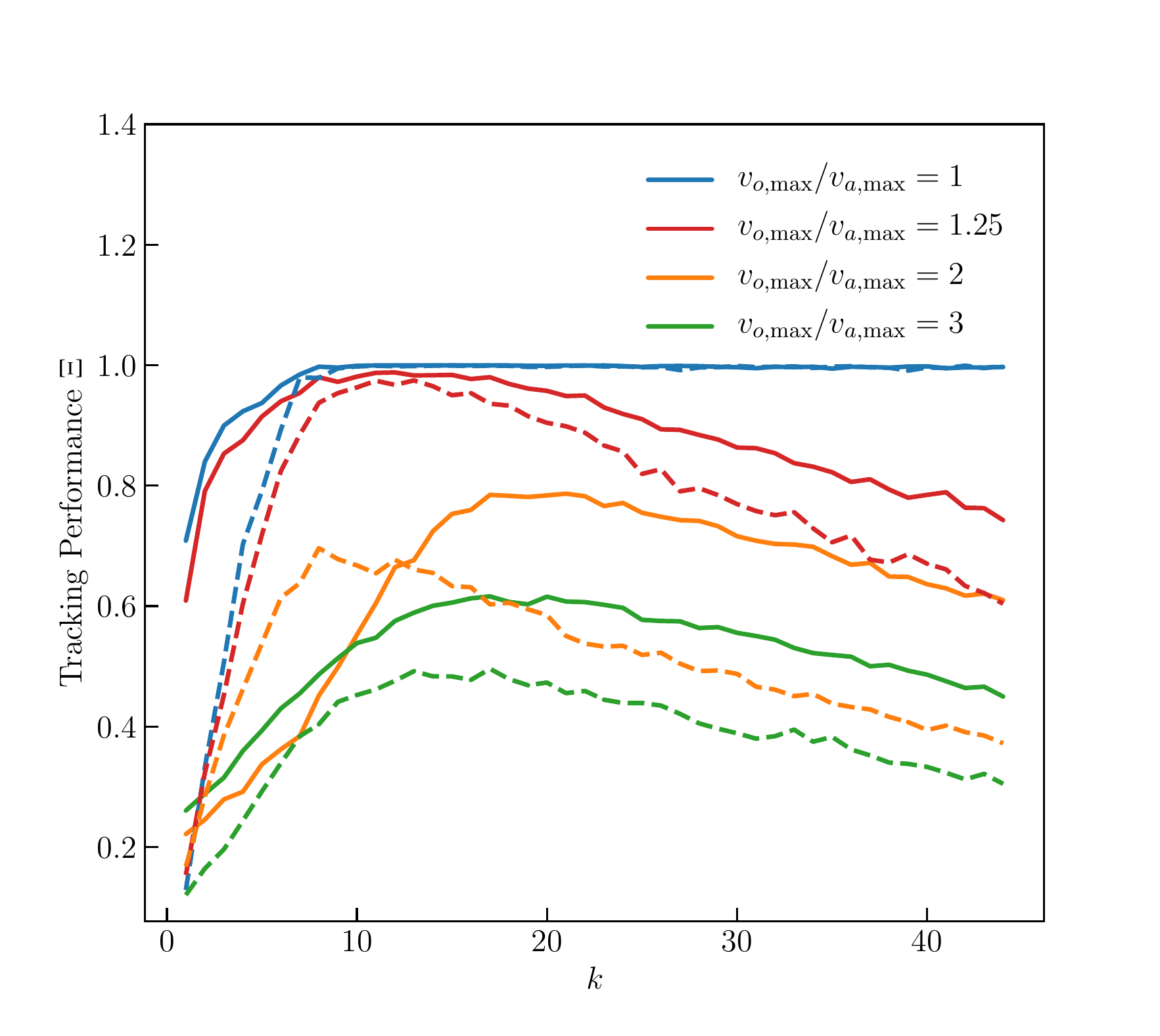}
    \vspace{-3ex}
    \caption{Non-evasive target (solid line) and evasive target (dashed line) tracking performance with different $v_{o,\text{max}}/v_{a,\text{max}}$ ratios (maximum agent's velocity $v_{a,\text{max}}=0.1$ and memory length of $t_{\text{mem}}=20$).}
    \label{fig:single_evasive_k}
\end{figure}

While tracking both evasive and non-evasive targets, increases in target velocity caused a decrease in the system's tracking performance as seen in Fig.~\ref{fig:single_evasive_k}. This was expected since the swarming agents' velocities were not increased to match that of the target and increasing a target's velocity allows it to more easily outrun its pursuers. Unsurprisingly, a reduction in tracking performance was also seen when targets used an evasive movement policy instead of the non-evasive one as it gave the targets the ability to better avoid its pursuing agents.

\begin{figure}[htbp]
    \centering
    \includegraphics[width=0.45\textwidth]{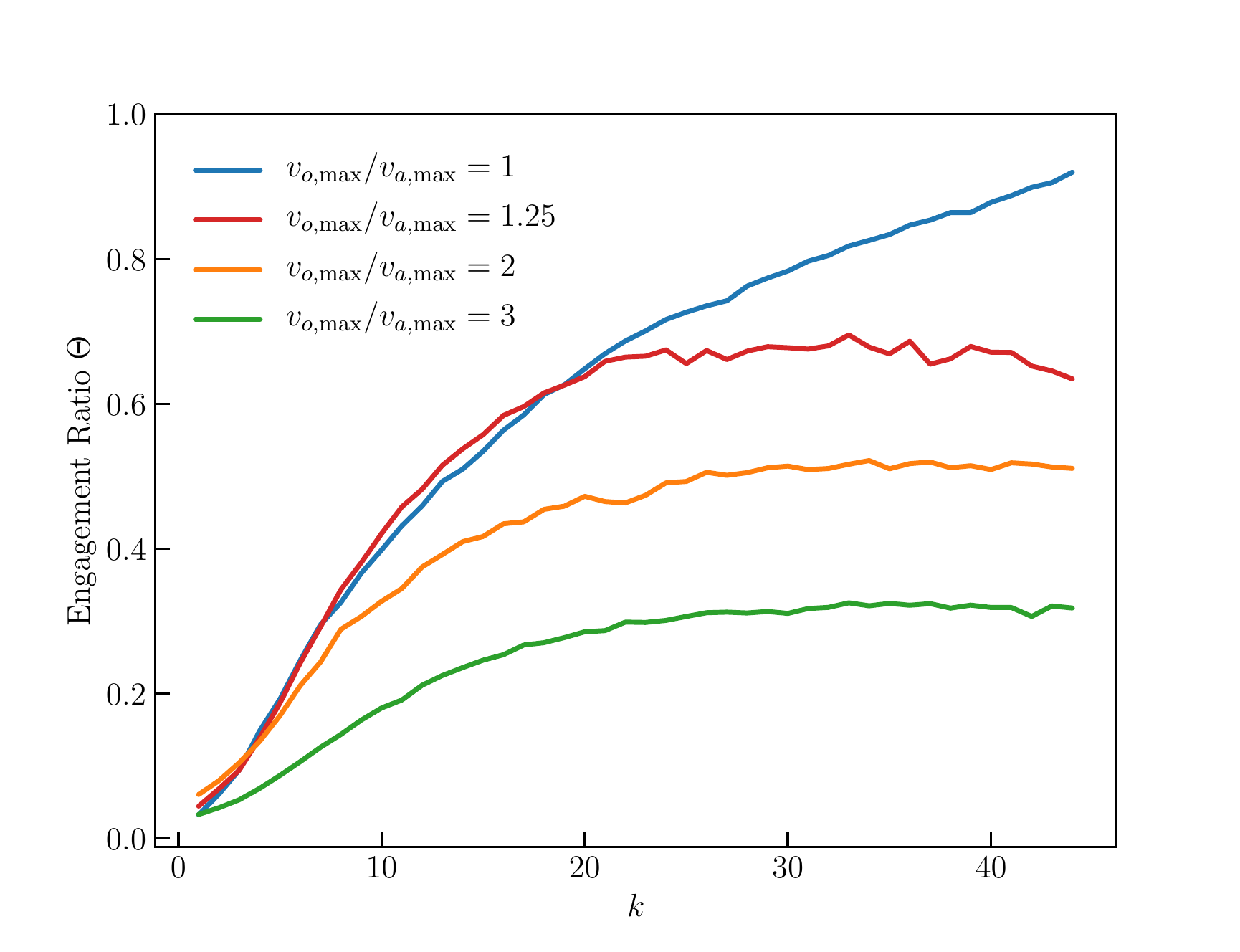}
    \vspace{-3ex}
    \caption{Swarm system engagement with a single evasive target traveling at different speeds (maximum agent's velocity $v_{a,\text{max}}=0.1$ and memory length of $t_{\text{mem}}=20$).}
    \label{fig:single_evasive_k_engage}
\end{figure}

From Fig.~\ref{fig:single_evasive_k}, it can also be observed that there exists an optimal degree of connectivity, $k=k^*$, at which the tracking performance is maximized and is also consistent with our previous work \citep{Kwa2020a, Kwa2020b} as well as the work carried out by \cite{Hamann2018}. However, this optimum only appears when the target travels faster than the individual agents. This is because at low target velocities, the swarm need not regulate its level of exploration and exploitation to effectively track the target; the swarm can track the target through performing purely exploitative actions. In contrast, when the target travels faster than the agents, the swarm needs to engage in more exploratory actions within the search space to reacquire the target's location after being outrun. \cite{Kwa2020a, Kwa2020b} have shown that a swarm tends to favor exploitation at high levels of connectivity and exploration at lower connectivity levels. This is confirmed by Fig.~\ref{fig:single_evasive_k_engage} that shows higher system engagement at higher levels of connectivity. As such, when operating at the optimal degree of network connectivity, it can be said that the system is able to carry out a good balance of both exploratory and exploitative actions, improving its target tracking capabilities.

The presence of an optimal $k$ is also apparent in non-evasive target tracking. However, $k^*$ for non-evasive target tracking tends to be higher than that for evasive target tracking. This is because an evasive target makes its movements to avoid contact with a pursuing agent, thereby requiring more exploration from the MRS to track the target. Exploration is favored at lower degrees of connectivity, resulting in a lower $k^*$ and associating evasive target tracking with a different exploration and exploitation balance.

Figure~\ref{fig:single_evasive_k_engage} also reveals an upper limit in the engagement ratio of the system in both evasive and non-evasive targets tracking. At higher levels of connectivity, the swarm loses track of the target easily due to a high portion of agents aggregating around and exploiting the target's location. Eventually, when the agents are outrun by the target, they have to expand and explore the environment again to relocate the target. This suggests that the amount of exploitation that can be performed by the swarm is being limited by the lack of information gathered by the swarm about the target's location, further stressing the critical need to balance the amount of exploration and exploitation carried out by the swarm.

\subsubsection{Multiple Target Tracking}
\begin{figure}[htbp]
    \centering
    \includegraphics[width=0.45\textwidth]{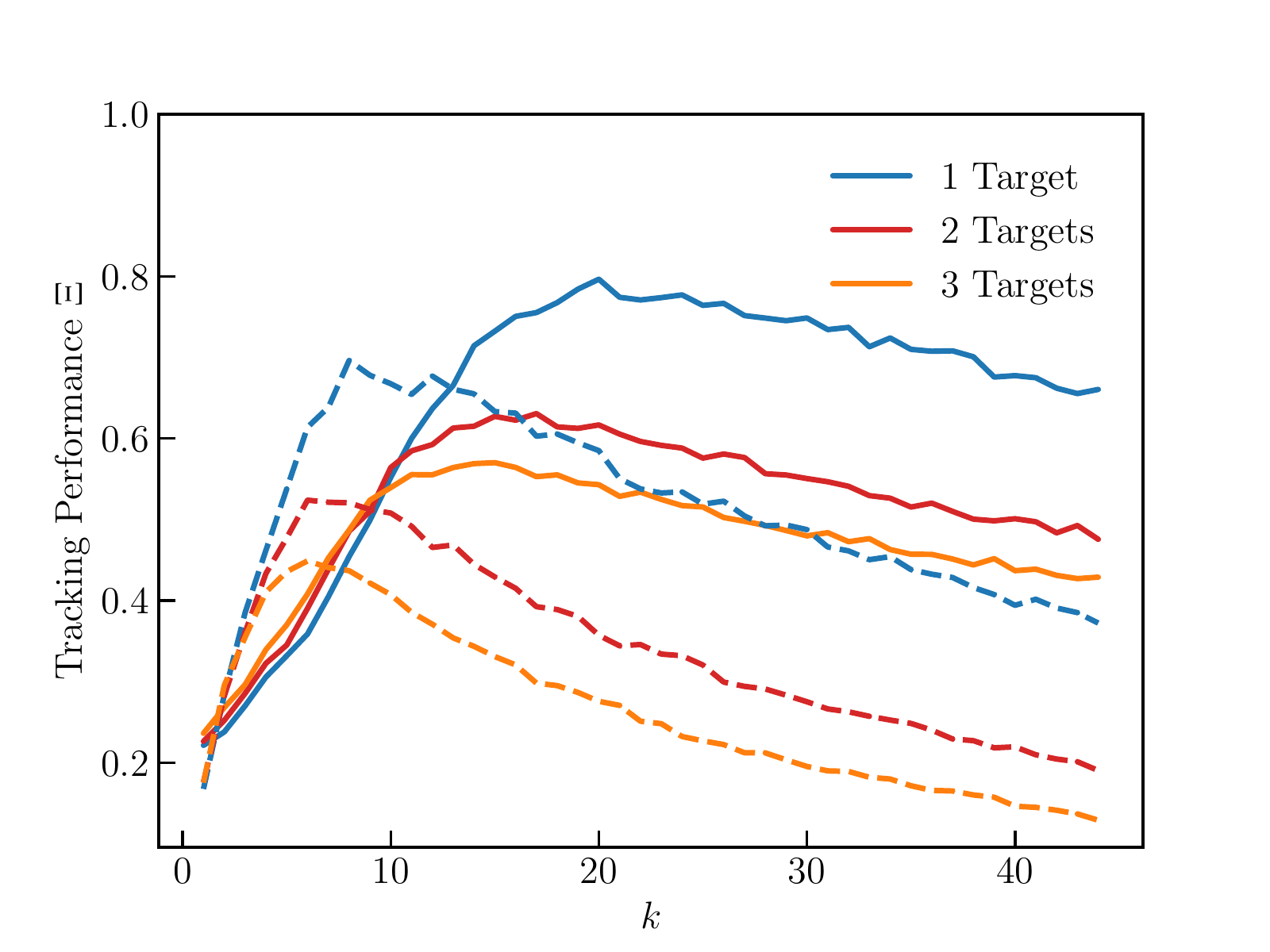}
        \vspace{-3ex}
    \caption{Multiple non-evasive (solid lines) and evasive (dashed lines) target tracking performance (maximum agent's velocity $v_{a,\text{max}}=0.1$, memory length of $t_{\text{mem}}=20$, and maximum targets' velocity $v_{o,\text{max}}=0.2$).}
    \label{fig:multiple_k}
\end{figure}

Similar to single target tracking, the swarm has a lower $k^*$ when tracking evasive targets compared to when tracking non-evasive targets (see Fig.~\ref{fig:multiple_k}). Again, this is due to the swarm requiring higher levels of exploration when pursuing evasive targets.
Also seen in Fig.~\ref{fig:multiple_k} is a reduction in $k^*$ when tracking multiple targets. This holds true for both evasive and non-evasive targets and is because more exploration is required to acquire information about the different targets' location. These higher levels of exploration are achieved at lower levels of connectivity. It is also harder to track multiple targets simultaneously. Therefore, increasing the number of targets present in the search space lowers the tracking performance of the swarm.

\label{sec:sim_mem}
\begin{figure}[ht!]
    \centering
    \includegraphics[width=0.45\textwidth]{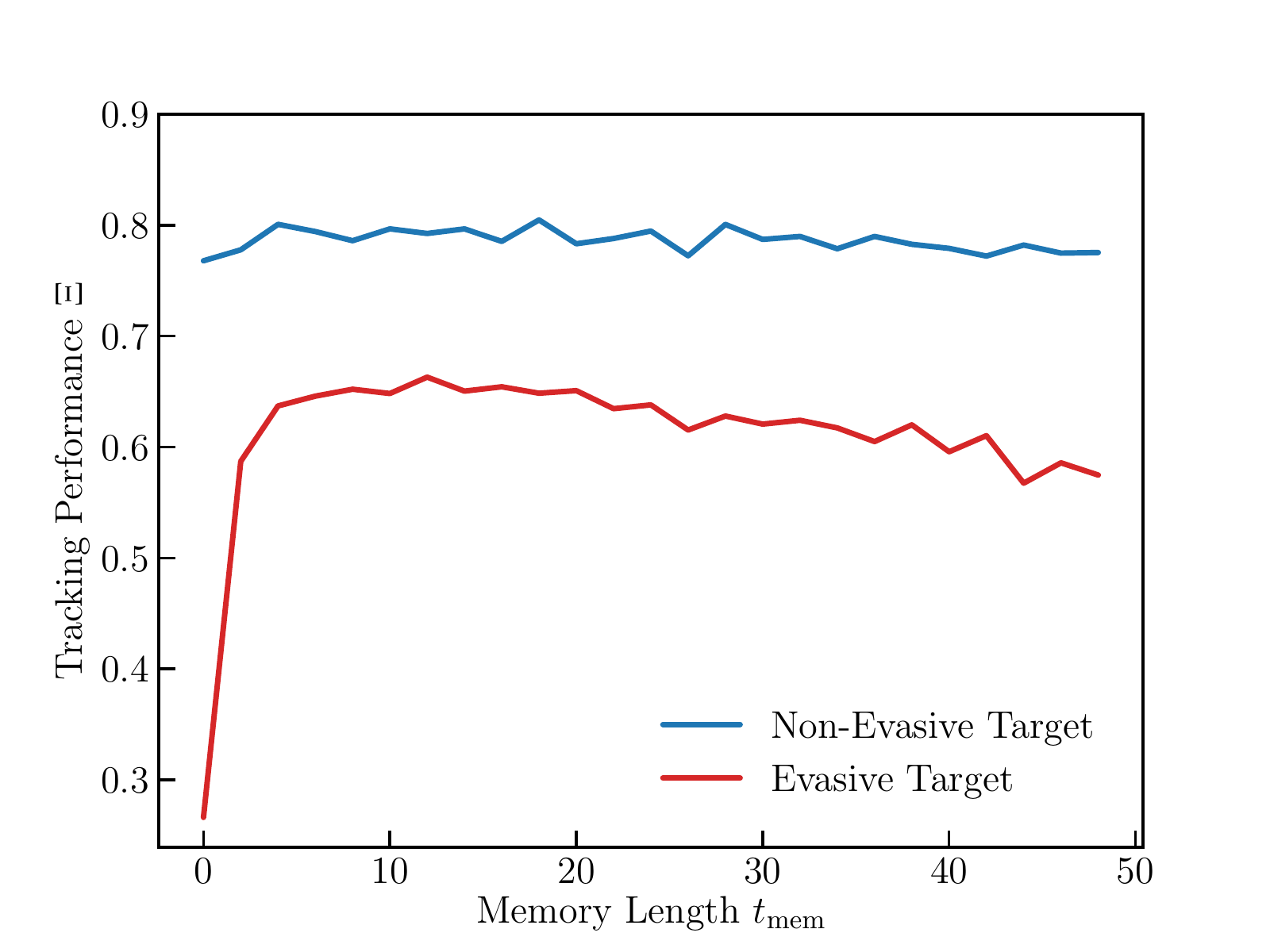}
    \caption{Single evasive and non-evasive target tracking performance with different implemented agent-based memory lengths (target speed $v_{o,\text{max}}=0.2$). Tracking agents are connected in a dynamic $k=20$ network.}
    \label{fig:memory_comparison}
\end{figure}

\subsection{Impact of Varying Agent-Based Memory}
It has previously been assumed that the addition of agent-based memory will cause the swarm to exploit outdated information, reducing the system's overall tracking performance. However, as seen in Fig.~\ref{fig:memory_comparison}, the addition of moderate amounts of agent-based memory to the system only results in a small decrease an MRS's non-evasive target tracking performance.

\begin{figure}[ht!]
    \centering
    \includegraphics[width=0.45\textwidth]{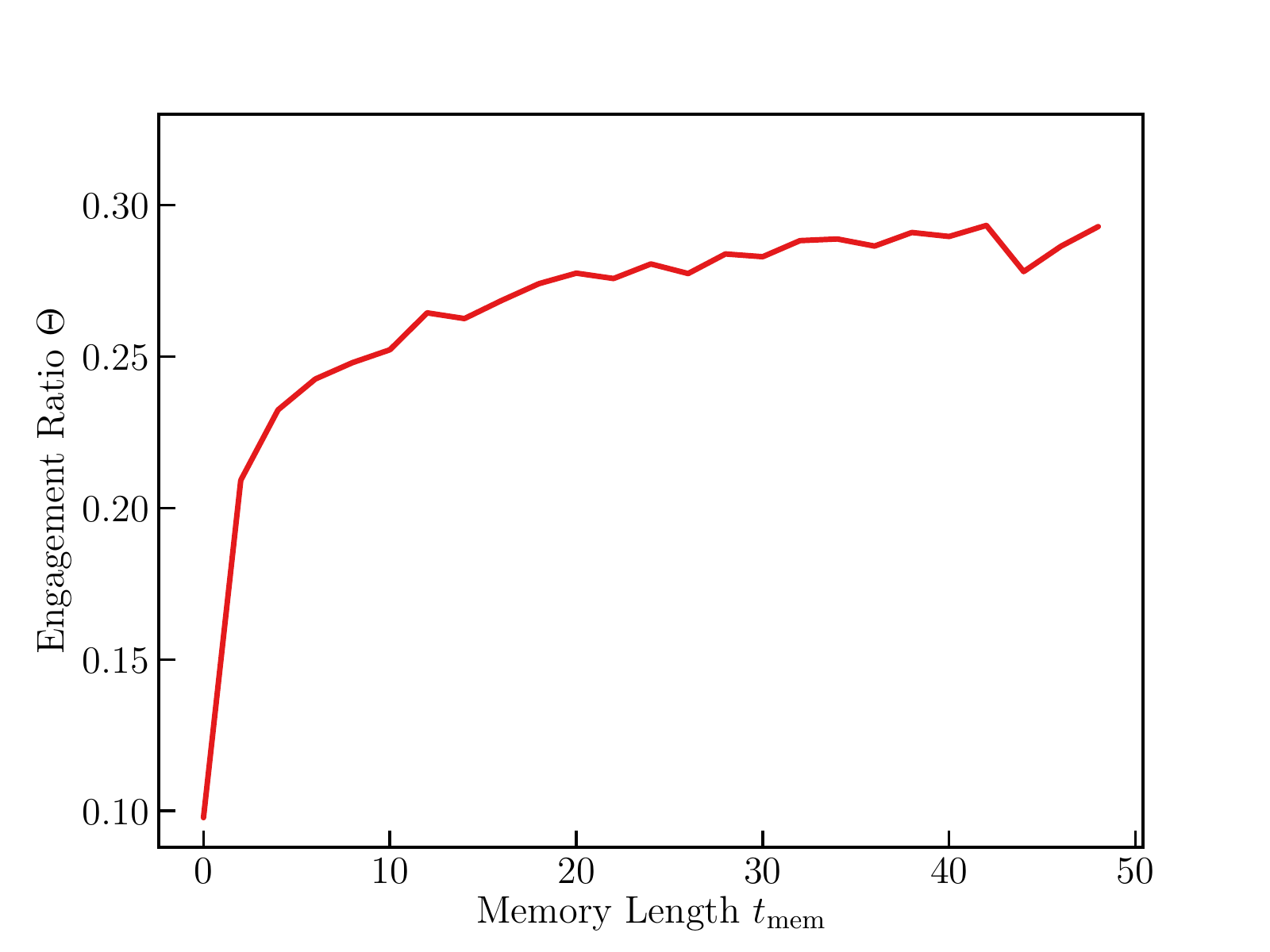}
    \caption{Evasive target system engagement with varying memory lengths. Target moved at a speed of $v_{o,\text{max}}=0.2$. Agents were connected in a dynamic $k=20$ network.}
    \label{fig:evasive_memory_engagement}
    \includegraphics[width=0.45\textwidth]{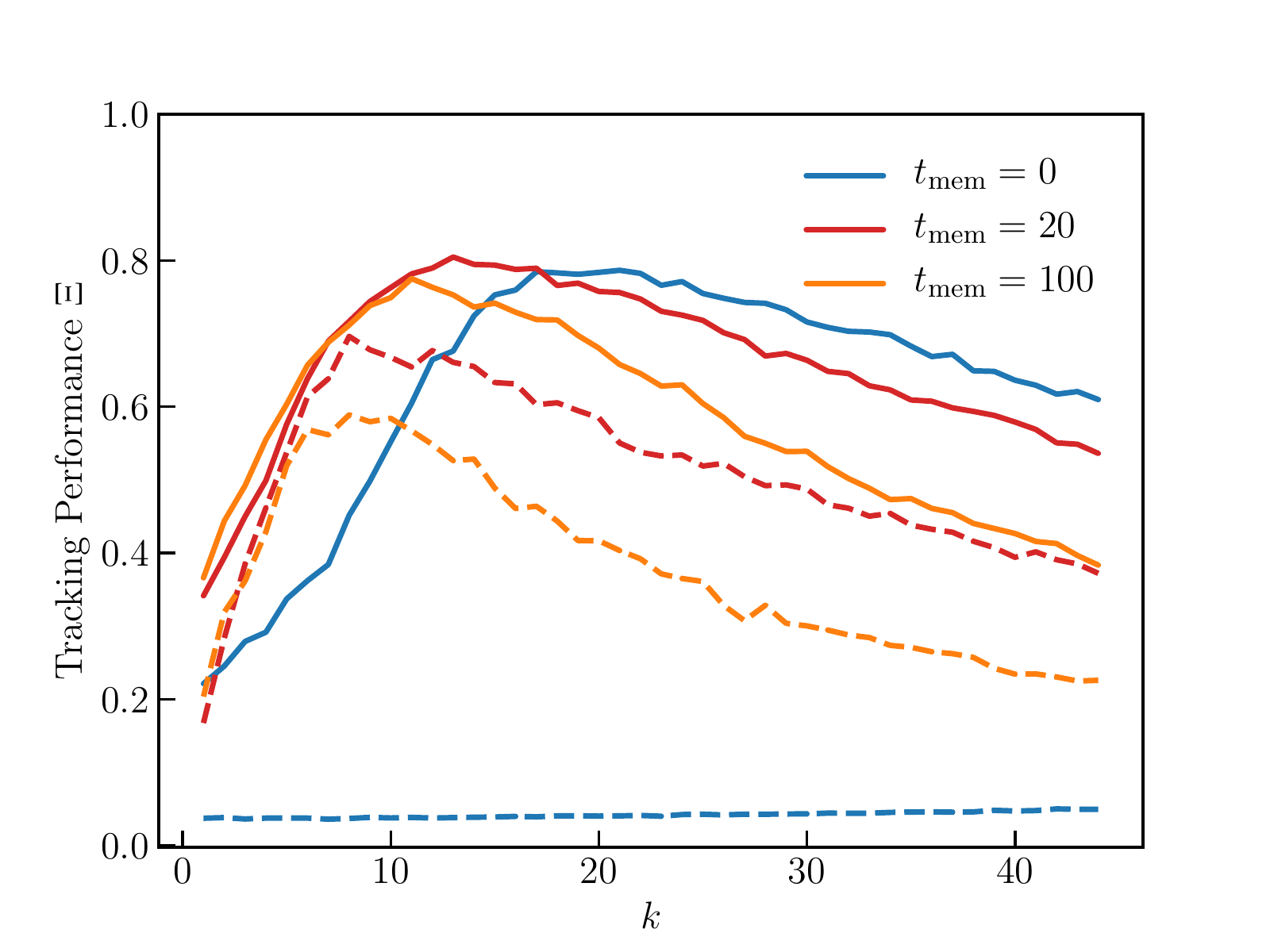}
    \caption{Single non-evasive (solid line) and an evasive (dashed line) target tracking performance while varying the length of the implemented agent-based memory and degree of network connectivity.}
    \label{fig:memory_optimal_k}
    \vspace{-2ex}
\end{figure}

In contrast, when tracking an evasive target, memory plays a major role in a swarm's ability to track the target. Without its use, the swarm is unable to track the target effectively. As demonstrated in Fig.~\ref{fig:memory_comparison}, the tracking performance rapidly improves when memory is introduced until an optimum memory length is reached before slowly degrading again. This is because the use of agent-based memory generates a persistent point of attraction based on a target's last known position, giving the agents the ability to aggregate at that point. While the swarm may periodically encounter the target without the use of memory, its agents will tend to expand until they reach their static equilibrium positions and will be unable to close in on the target due to the target's evasive maneuvers. This is illustrated in Fig.~\ref{fig:evasive_memory_engagement} where it can be seen that the swarm is unable to engage with the target with very low agent-based memory lengths.

\subsection{Balancing Network Connectivity and Memory}

When comparing Figs.~\ref{fig:single_evasive_k}~and~\ref{fig:memory_comparison}, it can be seen that after the optimal memory length has been attained, changing the degree of connectivity impacts the MRS's tracking performance more compared to the effects attained by altering the memory length. However, Fig.~\ref{fig:memory_optimal_k} shows that changing the length of the memory present in each agent also affects the optimal degree of connectivity required to maximize the tracking performance. It can be seen that increasing agents' memory length tends to decrease the optimal degree of connectivity, $k^*$. This is caused by the swarm's tendency to perform higher amounts of exploitation with longer memory lengths. This is congruent with the findings by \cite{Nauta2020a}, who showed that stronger memory effects favor more exploitative actions in a foraging scenario. Therefore, to maximize its tracking performance while using increased memory lengths, the swarm needs to compensate for this increase in exploitation by performing more exploratory actions, effectively resulting in a reduced $k^*$ value.

\begin{figure}[h]
\centering
    \includegraphics[width=0.49\textwidth]{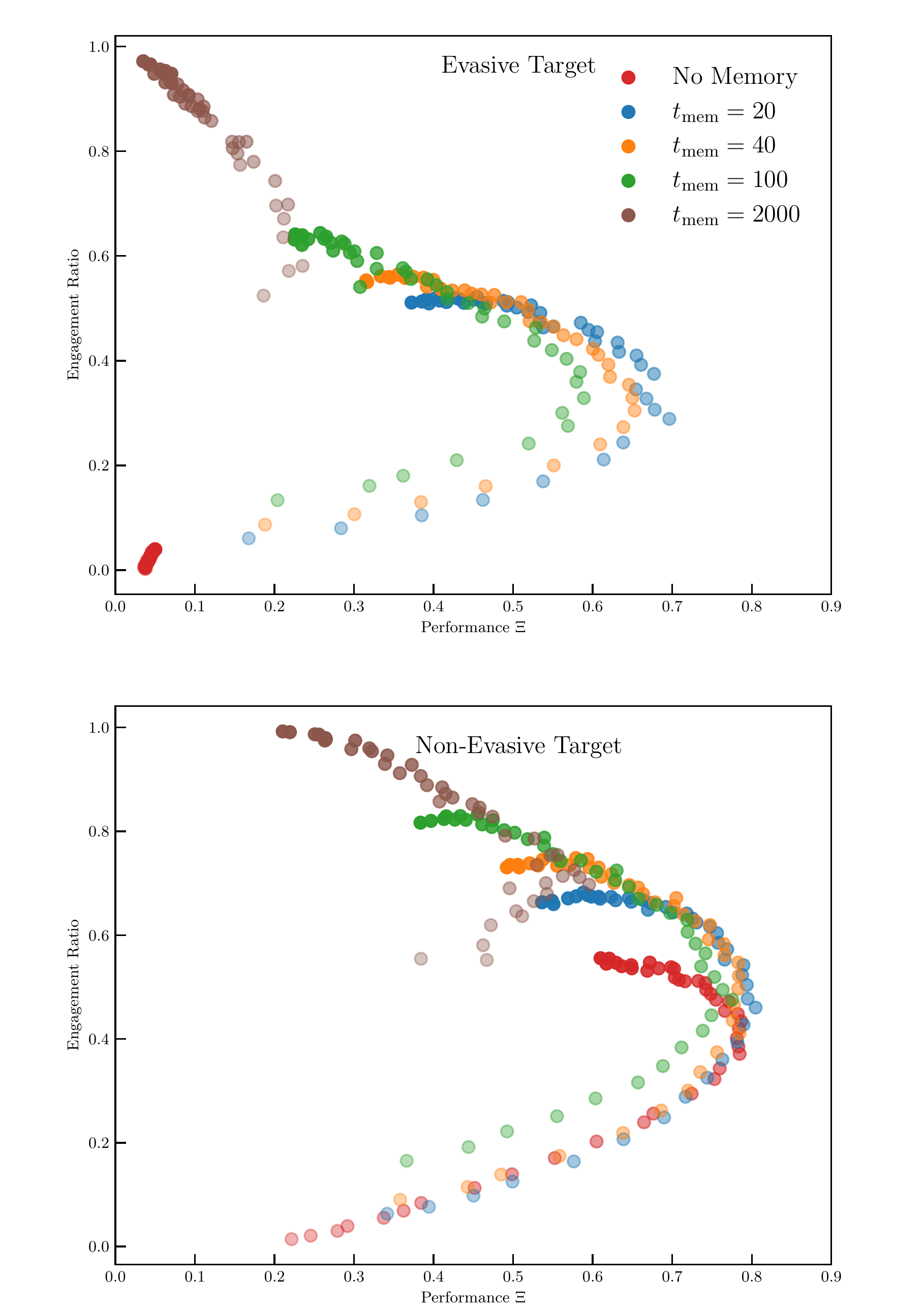}
        \vspace{-2ex}
    \caption{Engagement-Tracking plots of a swarm with different $t_\text{mem}$ and $k$ tracking an evasive~(top) and a non-evasive~(bottom) target. Darker shaded points indicate swarms using networks with higher values of $k$.}
    \label{fig:engagement_tracking}
\end{figure}

The results obtained point to an optimal balance between exploration and exploitation carried out by the swarm to maximize its tracking response. This is demonstrated in Fig.~\ref{fig:engagement_tracking}, which shows a clear optimum level of engagement that maximizes the MRS's tracking performance. The plots also show that the ideal engagement ratio maximizing evasive target tracking performance is lower than that for a non-evasive target. Not only do these results highlight the presence of the optimal balance between exploration and exploitation, but they also reveal the important fact that this optimum varies based on the task presented to the swarm.

Furthermore, Fig.~\ref{fig:engagement_tracking} shows that excessive memory lengths ($M = 2000$) degrade the tracking performance of the swarm when pursuing both evasive and non-evasive targets. The plot also highlights the necessity of the inclusion of agent-based memory when tracking evasive targets as the system without memory is outperformed by all the other systems.

\section{Swarm Robotic Experiments}
\begin{figure*}[htbp]
    \centering
    \includegraphics[width=0.75\textwidth]{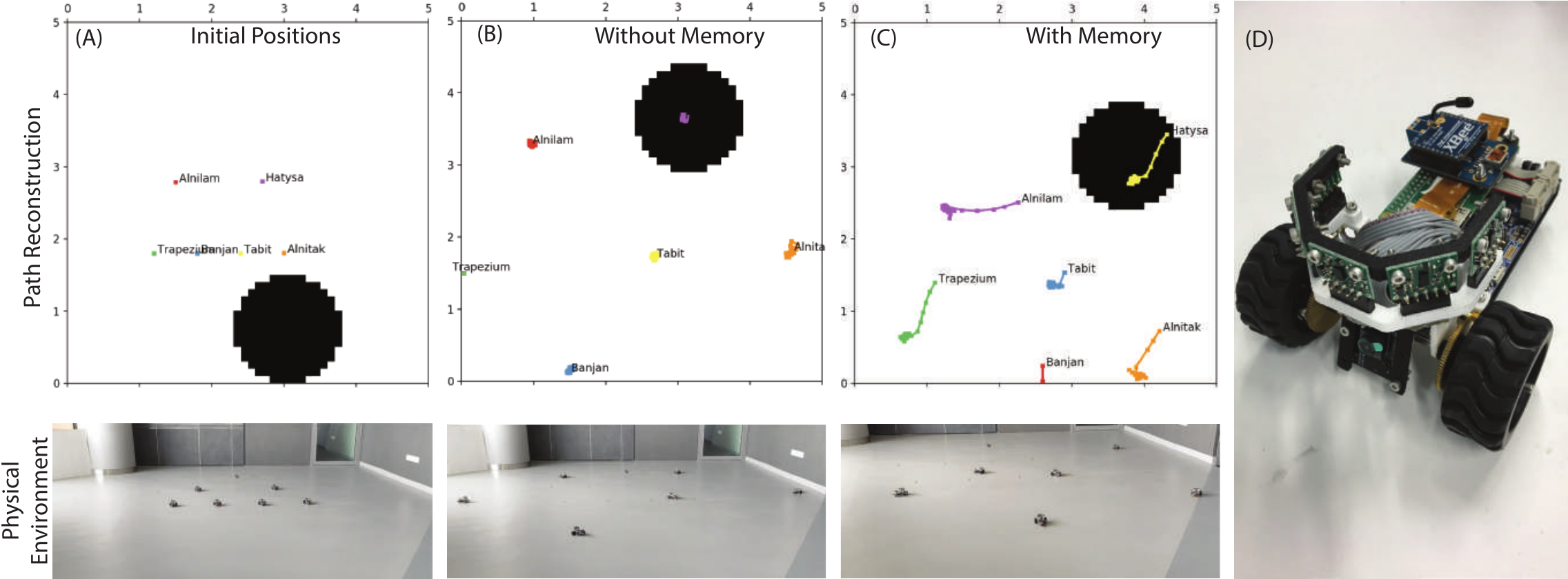}
        \vspace{-2ex}
    \caption{(Top) Reconstruction of target and robot positions and robot paths as observed through our monitoring system. (Bottom) Physical experiments using a connectivity of $k=5$ and adaptive repulsion. (A) Robots initialized at the start of the experiments. (B) Robots without agent-based memory unable to respond to a target's presence. (C) Robots with agent-based memory moving towards target after being found. (D) ORION robotic unit~\citep{Kit2019}.}
    \label{fig:experiment}
\end{figure*}
%
To validate the obtained simulation results, we performed series of experiments using 6 in-house developed differential-drive land robots with a maximum speed of $10~\text{cm}/\text{s}$. Although this swarm robotic system was originally developed for the mapping of unknown environments~(see \cite{Kit2019} for technical details), it has also been used in the tracking of non-evasive targets~\citep{Kwa2020a}.

At the start of all tests, 6 of the robots were arranged in the middle of an open search area. These robots were tasked with tracking a virtual evasive target, traveling at an average speed of $5~\text{cm}/\text{s}$, following the same evasive movement policy previously described with $\rho = 0.75~\text{m}$. The target's location is determined by a central computer that calculates its next position based on the current locations of all of robots. To ensure that no gradient descent methods are used, only the units within the target's radius are given information about the target's location. It must be emphasized that besides this communication of target locations to the robots and robot positions to the computer, there was no central controller facilitating the coordination of the robots' movements. It should also be noted that while the target traveled at a slower speed than the robots, the target has infinite maneuverability compared to the finite maneuverability of the robots. As such, even if the individual robots can move faster than the target in a straight line, they effectively respond much slower due to their low maneuverability. It is also worth mentioning that these experiments were carried out in an environment free of obstacles to highlight the effects of agent-based memory. Should this experiment be rerun in a more complex environment with obstacles, the robot units are equipped with LIDAR sensors, allowing them to perform collision avoidance.

All experiments were run with an all-to-all connectivity ($k=5$), in an attempt to make the response of the swarm more apparent. Two sets of tests were run, one with agent-based memory length of $M = 15~\text{s}$, and another without any. Each set of tests was comprised of 10 runs, with each run lasting for approximately 2 minutes to minimize the directional drift experienced by the robots and allow for the assumption that the robots are able to accurately self-localize. 



The robots' locations are communicated to a central computer to enable the reconstruction and visualization of their paths and responses to the presence of a target seen in Fig.~\ref{fig:experiment}. The figure shows that without memory, the robots were unable to respond to the presence of the target and tended to remain in place even when the target had been encountered. In contrast, with agent-based memory, when an agent encountered the target, the swarm as a whole was able to respond and the units turned to move towards and aid in the tracking of the target. The improved response of swarm system to the presence of the target resulted in a significant increase in tracking performance $(p < 0.01)$, with the swarm being better able to track the evasive target for a longer period of time (Table~\ref{tab:results}). This is similar to the results obtained in the simulations, validating our hypothesis that agent-based memory is required in the tracking of a fast-moving evasive target.

While the physical experiments performed confirmed the necessity of the implementation of agent-based memory, the quantitative results obtained from the physical experiments did not replicate those obtained from the simulations. These discrepancies can be attributed to three reasons: (1) the small number of robots used for the simulations, (2) the short experiment duration relative to that of the simulations, and (3) the physical tests and the simulations were not performed at the same scale and density. Also as demonstrated by \cite{Czirok1997}, statistical characterization of a multi-agent system's order becomes more difficult at low swarm densities. This makes it harder to predict the behavior of a system with low density within simulations, which could account for some of the discrepancies between the simulation and physical experiment results.

\begin{table}[htbp]
    \caption{Swarm tracking performance results with and without agent-based memory obtained from the robotic units.}
    \vspace{-3ex}
    \begin{center}
        \begin{tabular}{lcccl}
        \label{tab:results}
                 & \multicolumn{2}{c}{Without Memory} & \multicolumn{2}{c}{With Memory} \\\hline
                 & 39.0             & 31.8            & 48.8           & 63.8           \\
                 & 47.1             & 39.9            & 46.6           & 44.0           \\
                 & 37.6             & 40.5            & 64.4           & 37.0           \\
                 & 32.9             & 37.1            & 40.1           & 58.4           \\
                 & 30.8             & 35.1            & 48.1           & 42.8           \\ \hline
        Averaged & \multicolumn{2}{c}{$37.2 \pm 4.6$}           & \multicolumn{2}{c}{$49.4 \pm 9.2$}        \\ \hline
        \end{tabular}
    \end{center}
\end{table}

\section{Discussion}
The application of swarming MRS to a dynamic target search and tracking task is a very challenging problem. This is especially the case when multiple evasive targets capable of traveling faster than the individual agents are being considered. Such cases are not adequately studied as it has been assumed that a fast-moving evasive target will always be able to outmaneuver its pursuers, rendering the tracking task impossible and therefore trivial.

In this work, we show that despite having to track multiple fast-moving evasive targets, it is possible to employ limited perception swarming MRS to accomplish the task through the use of a decentralized swarming strategy. The presented strategy allows the tuning of a swarm's exploration and exploitation dynamics through the use of an adaptive inter-agent repulsion behavior, the degree of connectivity of the interconnecting $k$-nearest neighbor network, and the length of memory present in all agents. The swarm's EED shifts in favor exploration by reducing the degree of connectivity or by reducing the agents' memory lengths and vice versa to favor information exploitation. This strategy was tested in simulations where we quantified the swarm's overall tracking performance and its EED. The former was performed by counting the number of time-steps the targets had been tracked and the later was done by calculating the swarm's engagement ratio, the average proportion of agents attempting to track a target during the entire simulation.

Through tuning the swarm's EED, an optimum balance between the level of exploration and exploitation was found, occurring only when the swarm tracks targets moving faster than its component agents. This balance tilts in favor of performing higher amounts of exploration when attempting to track multiple evasive targets, due to the swarm's need to acquire information of the targets' locations. Similarly, when tracking a smaller number of non-evasive targets, better tracking performance will be obtained if a swarm performs higher amounts of exploitative actions.

A limitation of this work is that our MRS operates in an unobstructed environment. However, as this work focuses on the highly dynamic problem of tracking a fast-moving target and the importance of memory in the performance of such tasks, more complex environments were not used. Should the MRS be implemented within an environment with obstacles, the simulations can be modified to utilize obstacle avoidance algorithms, taking advantage of our robotic test platform's LIDAR sensors.

Finally, physical experiments were carried out with a swarm of 6 robots. While the quantitative results did not replicate those obtained in the simulations, we demonstrated a significant improvement of the swarm's evasive target tracking performance when equipped with agent-based memory compared to an MRS without. This is in agreement with our results, highlighting the importance of memory in the search and tracking of a fast-moving evasive target.

\section{Acknowledgements}
This research is funded by Thales Solutions Asia Pte Ltd, under the Singapore Economic Development Board IPP.

\footnotesize

\end{document}